\documentclass[10pt]{article}

\usepackage{amsmath}
\usepackage{color}
\usepackage[colorlinks,citecolor=blue,urlcolor=blue,linkcolor=blue]{hyperref}
\usepackage[utf8]{inputenc}
\usepackage{graphicx}
\usepackage{float}
\usepackage{bm}
\usepackage{braket}
\usepackage{enumerate}
\usepackage{authblk}

\numberwithin{equation}{section}

\newcommand{\Hb}{\overline{\mathrm{H}}}
\newcommand{\gb}{\overline{g}}
\newcommand{\dd}{\mathrm{d}}
\newcommand{\Ai}{\text{Ai}}
\newcommand{\hor}{\mathrm{hor}}
\newcommand{\tot}{\mathrm{tot}}

\begin{document}

\title{Shaping the distribution of vertical velocities of antihydrogen in GBAR}
\date{}

\author[1]{G.~Dufour}
\author[2]{P.~Debu}
\author[1]{A.~Lambrecht}
\author[3]{V.V.~Nesvizhevsky}
\author[1]{S.~Reynaud}
\author[4]{A.Yu.~Voronin}

\affil[1]{Laboratoire Kastler-Brossel, CNRS, ENS, UPMC, Campus Jussieu, F-75252 Paris, France}
\affil[2]{Institut de Recherche sur les lois Fondamentales de l'Univers, CEA-Saclay, F-91191 Gif sur Yvette, France }
\affil[3]{Institut Max von Laue - Paul Langevin, 6 rue Jules Horowitz, F-38042, Grenoble, France }
\affil[4]{P.N. Lebedev Physical Institute, 53 Leninsky prospect, Ru-117924 Moscow, Russia }

\maketitle

\begin{abstract}
GBAR is a project aiming at measuring the free fall acceleration of
gravity for antimatter, namely antihydrogen atoms ($\Hb$). Precision
of this timing experiment depends crucially on the dispersion of
initial vertical velocities of the atoms as well as on the reliable
control of their distribution. We propose to use a new method for
shaping the distribution of vertical velocities of $\Hb$, which
improves these factors simultaneously. The method is based on
quantum reflection of elastically and specularly bouncing $\Hb$ with
small initial vertical velocity on a bottom mirror disk, and
absorption of atoms with large initial vertical velocities on a top
rough disk. We estimate statistical and systematic uncertainties,
and show that the accuracy for measuring the free fall acceleration
$\gb$ of $\Hb$ could be pushed below $10^{-3}$ under realistic
experimental conditions.

\vspace{0.3cm}



\end{abstract}

\section{Introduction}

Gravitational properties of antimatter have never been measured
directly. A promising experimental method to do so consists in
producing sufficiently cold antihydrogen atoms ($\Hb$) and timing
their free fall in the Earth's gravity field. This approach is being
pursued by AEGIS \cite{kellerbauer2008}, ATHENA-ALPHA \cite{alpha},
ATRAP \cite{atrap} and GBAR \cite{gbar} collaborations.

In order to get the highest accuracy for measuring the free fall
acceleration $\gb$ of $\Hb$, one has to cool atoms down to low
temperatures and to measure, or at least to deduce from design and
calculations, the initial velocity distribution. We discuss here the
method proposed by Walz and H\"{a}nsch \cite{walz2004} which is used
in the GBAR project to reach very low temperatures~: ${\Hb}^+$ ions
are trapped and cooled down to the lowest quantum state in a Paul
trap, and  $\Hb$ is then produced by photo-detaching the excess
positron. The photo-detachment pulse is the START signal for the
free fall timing measurement, while the STOP signal is provided by
the annihilation of $\Hb$ atoms on a detection plate placed at a
height $H$ below the center of the ion trap.

Precision of this measurement depends crucially on the dispersion of vertical
velocities before the free fall, which corresponds to the residual kinetic energy of the atoms after the cooling process.
The aim of the present paper is to propose a new filtering method to further reduce
the initial distribution of vertical velocities  and
thus improve the accuracy in the measurement of $\gb$.

In section \ref{spreading} we justify our choice of characteristic
values for the spatial localization of the initial atomic cloud by
considering the spreading of the freely-falling wave-packet of $\Hb$
in the gravitational field. We describe in section \ref{shaping} the
new method for shaping vertical velocities of $\Hb$ in the
quasi-classical approximation, and show in section \ref{uncertainty}
that the improvement of accuracy due to the velocity selection
overcomes the degradation associated with the decrease of the
statistics. We then present in section \ref{qmech} a
quantum-mechanical description of the experiment in order to
validate the quasi-classical estimations of the preceding sections.
In section \ref{systematics} we list possible systematic effects and
show that they scale down compared to those in the case of
unrestricted free fall of $\Hb$. We then deduce the accuracy which
could be reached on the measurement of $\gb$ under realistic
experimental conditions.

We neglect throughout this paper systematic effects
related to the energy-dependent probability of quantum reflection of
$\Hb$ from the detection plate \cite{dufour2013}.
The atomic recoil in the photo-detachment process induces
 an additional velocity dispersion which is discussed in the last section on systematic effects.

\section{Spreading of a freely-falling wave-packet}
\label{spreading}

In the simplified description presented in the introduction, the
initial distribution at time  $t=0$ is the lowest quantum state in
the Paul trap. This corresponds to a Gaussian wave-packet with
vertical velocity dispersion $\upsilon$ and vertical position
dispersion $\zeta$ reaching the minimum in the Heisenberg
uncertainty relation:
\begin{align}\label{heisenberg}
m\upsilon \zeta =\frac{\hbar }{2}
\end{align}
where $\hbar$ is the reduced Planck constant and $m$ the inertial
mass of $\Hb$.

After their release from the trap at time  $t=0$, atoms start
falling freely in the Earth's gravity field until they reach the
detection plate placed at a height $H$ below the center of the trap.
The time of fall is measured as the delay $t$ from their release to
their annihilation on the detection plate. The acceleration of
gravity $\gb$ for antihydrogen is then deduced from the distribution
of these fall times. This acceleration $\gb$ for antihydrogen is
related to the analog quantity $g$ defined for hydrogen by
$\gb=Mg/m$, where $M$ is the gravitational mass of $\Hb$.

We now discuss the distribution of free fall times, assuming for
simplicity that this distribution is determined by initial
dispersions of vertical velocity and position (other sources of
uncertainty negligible). If the initial quantum state is poorly
localized (large values of $\zeta$) then the spread of the fall
times is too large because of the initial dispersion of height. In
the opposite case where the wave-packet is too localized (small
values of $\zeta$) then the spread of the fall times is too large
because of initial dispersion of vertical velocity. An optimum
localization of the initial quantum state should be found as a
compromise between these two limiting cases.

As the variations of position and velocity are uncorrelated in the
initial wave-packet, a classical calculation gives the relative
spread $\Delta t$ of the free fall times arising from both effects:
\begin{align}
\frac{\Delta t}{t_H}=&
\sqrt{{\left(\frac{\zeta}{2H}\right)}^{2}+{\left(\frac{\upsilon}{v_H}\right)}^2} \label{deltat}\\
=&\sqrt{{\left(\frac{\zeta}{2H}\right)}^{2}+{\left(\frac{\hbar }{2m
v_H\zeta }\right)}^{2}}~. \label{deltatheisenberg}
\end{align}
The second of these relations uses \eqref{heisenberg} while the
first one is valid even when $\upsilon$ and $\zeta$ do not reach the
minimum in Heisenberg uncertainty relation. We have defined
$t_H=\sqrt{2H/\gb}$ and $v_H=\sqrt{2\gb H}$ as the free fall time
and velocity for a free fall height $H$ with zero initial velocity.
The optimum size of the initial state, which minimizes $\Delta t$ in
\eqref{deltatheisenberg}, is:
\begin{align}
{\zeta}_{{opt}}=\sqrt{\frac{{\hbar H}}{mv_H}}~.
\end{align}
It leads to an optimum resolution for the free fall measurement:
\begin{align}
{\left(\frac{\Delta t}{t_H}\right)}_{{opt}}=\sqrt{\frac{\hbar
}{2mv_HH}}~. \label{optimumresol}
\end{align}
The larger the product $mv_HH$ with respect to $\hbar /2$, the
better this optimal resolution is.

Better precisions are also obtained by increasing the fall height
with the characteristics of the trap kept fixed. However, the setup
size is limited by practical arguments involving price and space
considerations. Note that equation \eqref{optimumresol} is
translated in an uncertainty twice larger on the acceleration of
gravity
\begin{align}
\frac{\Delta\gb}{\gb}=2\frac{\Delta t}{t}
\end{align}
in the simple derivation presented here (a detailed analysis based
on Monte-Carlo simulations is given in \cite{gbar}).

With the typical numbers used for the design \cite{gbar} of the GBAR
experiment ($H=0.3$~m so that $v_H\approx2.4$~m/s if $\gb\approx
g$), one obtains $\zeta _{opt}\approx88$~$\mu$m and $\left(\Delta
t/t_H\right)_{opt} \approx2.1{\times}{10}^{-4}$. If this optimum
operation could be experimentally realized, the accuracy would reach
$\left(\Delta\gb/\gb\right)_{opt} \approx 4.2{\times}{10}^{-4}$ for
each detection of an annihilation event. With a total number of
events $N_{\tot } \approx2.6{\times}{10}^{4}$, calculated for a
typical measuring time of 1 month and an average production rate of
1 ultracold $\Hb$ atoms per period of 100~s, this would lead to the
resolution after one month:
\begin{align}
\left( \frac{\Delta \gb}{\gb \sqrt{N_{\tot }}} \right)_{opt}=
\sqrt{\frac{2\hbar}{m v_H H N_{\tot }}} \approx 2.6 \times 10^{-6}~.
\end{align}
We have assumed there were no large systematic effect.

However, the size of the initial cloud used in the design of the
GBAR experiment is far from this optimum.
The Paul trap is characterized by its oscillation frequency $\omega$ which fixes the velocity and position dispersions in the ground state:
\begin{align}
\zeta =\sqrt{\hbar /2m\omega } \quad,\quad \upsilon =\sqrt{\hbar
\omega /2m}~. \label{dispersions}
\end{align}
The mean kinetic energy in the ground state is then $m \upsilon^2/2=\hbar \omega/4$. 
Therefore the range of trap frequencies that can be used is limited by the residual kinetic energy of the atoms after cooling.

In GBAR, the considered frequency range is 0.1~MHz~$< \omega /2\pi <$~1~MHz, so that one gets
0.22~$\mu$m~$>\zeta>$~0.07~$\mu$m  and 0.14~m/s~$<\upsilon <$~0.44~m/s. This means that
the initial cloud is smaller than the optimum by about 3 orders of
magnitude. The resolution is thus limited by the dispersion of
initial velocity:
\begin{align}\label{accuracy}
\frac{\Delta\gb}{\gb\sqrt{N_{\tot}}}\approx\frac{2\upsilon}{v_H\sqrt{N_{\tot}}}~.
\end{align}

As it is not experimentally feasible to further cool down the ions to reach the
optimum size of the initial cloud, we propose in this paper to
select the initial vertical velocity of the atoms.
This will improve
the resolution after each annihilation event by a factor scaling as
the reduced velocity range $\Delta v /\upsilon $. The
statistics is reduced by a factor scaling as $\sqrt{N/N_{\tot
}}{\propto} \sqrt{\Delta v /\upsilon }$ (see equation \eqref{NNtot}) so that an overall
improvement is expected. Also systematic uncertainties will decrease
dramatically. The description of the shaping device and the
evaluation of its performance are discussed in more details in the
next sections.

\section{Shaping the distribution of vertical velocities of $\Hb$ in GBAR}
\label{shaping}

The current design for GBAR is a classical free-fall experiment
which aims at an accuracy of the order of 1\% \cite{gbar}. With a
quantum detection technique, one could get significantly higher
precision, in analogy to spectroscopy \cite{nesvizhevsky2002} or
interferometry \cite{nesvizhevsky2010a} of near-surface quantum
states \cite{nesvizhevsky2010b} of ultracold neutrons (UCNs)
\cite{ignatovich1990,golub1991}. However these techniques require
high energy resolution and sufficient statistics
\cite{voronin2011,voronin2012}. The method that we propose in this
paper is an intermediate step in this direction which is less
precise than the full quantum detection technique but allows for
better statistics and simpler design.

\begin{figure}
\centering
\includegraphics[width=8cm]{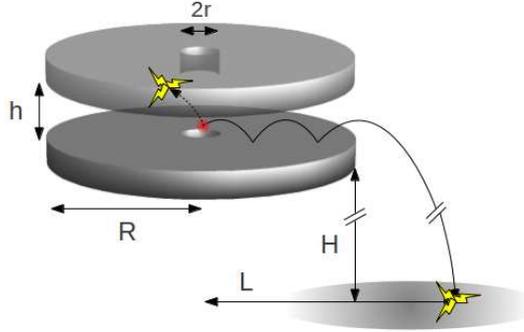}
\caption{\label{scheme} A scheme of principle of the proposed
shaping device: an $\Hb$ atom is released from the Paul trap
(central spot) and it bounces a few times on the mirror surface of
the bottom disk (arrows); if it scatters on the rough top surface,
it annihilates (lightnings); otherwise, it escapes from the aperture
between the two disks, and falls to the detection plate where it
annihilates (lightning on the detection plate).  $R$ is the radius
of the bottom and top disks,  $r$ is the radius of central openings
in the disks,  $h$ is the distance between the top surface of the
bottom disk and the bottom surface of the top disk,  $H$ is the
distance between the top of the detection plate and the top of the
bottom disk,  $L$ is the horizontal distance between the initial
spot and the detection point.}
\end{figure}

This method is analogous to the one used in the experiment on the
observation of gravitational quantum states of ultra-cold neutrons
\cite{nesvizhevsky2002,nesvizhevsky2003,nesvizhevsky2005,nesvizhevsky2000}.
The distribution of initial vertical velocities is shaped by
selecting the atoms passing through a shaping device consisting of
two disks. A scheme of principle of the shaping device where all useful quantities are defined is shown  in
figure \ref{scheme}. In the sequel of this section, a simple
analysis of the problem is presented in terms of quasi-classical
arguments, to be confirmed in the next sections. A more complete
quantum-mechanical description is also available in papers devoted
to ultra-cold neutrons
\cite{nesvizhevsky2005,voronin2006,meyerovich2006,adhikari2007,escobar2011}.

In the zone between the two disks, atoms with sufficiently small
vertical velocities bounce on the bottom mirror disk due to the high
efficiency of quantum reflection in the Casimir-Polder potential
\cite{dufour2013}. If the top surface of the mirror disk is flat,
smooth and horizontal, the horizontal velocity component as well as
the total energy of the vertical motion do not change and atoms thus
pass through the shaping device with high probability. This last
statement would be precisely valid for ideal quantum reflection from
the mirror surface; otherwise corresponding corrections have to be
taken into account (more discussions below). On the other hand,
atoms with large vertical velocities rise in the Earth's gravity
field to the height of the rough surface of the top disk and scatter
non-specularly on this surface. As this scattering mixes horizontal
and vertical velocity components, it leads to rapid loss of
scattered $\Hb$ through annihilation on the top or bottom disk.

A few remarks are useful at this point: 1) the shaping device has to
be coupled with the Paul trap (not shown on the figure); this point
is not discussed in this paper except for the role of the openings
left in the center of the disks for operating the Paul trap; note
that the disks may consist of several sectors not covering the
complete $2\pi$ horizontal angle in order to include the Paul trap
in the overall design; 2) annihilation events are supposed to be
detected with position-sensitive and time-resolving detectors; this
will allow one to account for the time spent in the shaping device
(see below); 3) due to the cylindrical symmetry of the device, all
atoms with small enough vertical velocity components and any value
and direction of the horizontal velocity component can pass through
it with high probability.


In order to describe the operation of the shaping device, we follow
possible classical trajectories of atoms from the initial point
where they are released to the points where they annihilate. As the
size of the initial spot (discussed in section \ref{spreading}) is
much smaller than any other characteristic size of the shaping
device, it plays no role in the following. We suppose the initial
spot of atoms to be placed at the height $H$ of the top surface of
the bottom disk (origin for altitude placed at the detection plate).
In a first step, we let the radius $r$ of the central opening tend
to zero and the radius $R$ of the disk tend to infinity.
Disregarding the losses due to imperfect quantum reflection on the
mirror disk, we obtain the fraction of atoms going through the
angular acceptance of the shaping device as:
\begin{align}\label{NNtot}
 \frac{N}{N_{\tot }}\approx\frac{\Delta v}{\upsilon}\sqrt{\frac{1}{2\pi }}
\end{align}
where $\upsilon$ is the standard deviation of the Gaussian
distribution of vertical velocities and $\Delta v$ the range of
vertical velocities fitting the aperture of the shaping device. With
the geometry sketched in figure \ref{scheme}, the latter corresponds to
atoms with vertical velocities $0<v<\Delta v$ with $\Delta v$
deduced from the energy needed to rise the height from $H$ to $H+h$
in the gravity field:
\begin{align}
 \Delta v=\sqrt{2\gb h}~.
 \label{deltavandh}
\end{align}

Note that the fraction of atoms going through the angular acceptance
of the shaping device changes as a function of the height of the
initial spot above the mirror disk as well as a function of the
radius $r$; therefore equation \eqref{NNtot} has to be modified for
other positions of the spot. Also equation \eqref{NNtot} has been
written in the limit of a good velocity selection  $\Delta v <
\upsilon $, which entails through \eqref{deltavandh} that $h$ has a
maximum value $h_{\max}$:
\begin{align}
    h<h_{\max}=\frac{{\upsilon }^{2}}{2\gb}~.
\end{align}
With the GBAR numbers considered above, the maximum value $h_{\max}$
lies in the interval 1-10~mm. If this condition is not obeyed,
equation \eqref{NNtot} has to be replaced by the appropriate
integral.

We now take into account the finite values of the radii of the
central openings $r$ and of the disks $R$. In order to do it
properly, we have to consider the shape of the angular distribution
of initial velocities. The operation of the Paul trap may indeed
require anisotropy to be introduced between horizontal and vertical
directions. This can be described by a ratio $\varepsilon $ between
frequencies of operation in horizontal and vertical directions
${\omega }_{\hor}={\varepsilon \omega }$ ($\omega $ is the frequency
already introduced for operation of the vertical trap). This ratio
should be in the interval $2<\varepsilon <4$ for a proper operation
of the Paul trap \cite{walther2012}. Using the same reasoning as in
the preceding section, we deduce that the horizontal dispersions are
\begin{align}\label{horizontal}
\upsilon_{\hor}=\upsilon \sqrt{\varepsilon }~,\quad
\zeta_{\hor}=\zeta /\sqrt{\varepsilon } ~.
\end{align}
where $\upsilon$ and $\zeta$ are the dispersions already introduced
for operation of the vertical trap.

We can now discuss the role of the finite radius $r$ of the central
openings. We want to avoid extra loss of statistics at the entrance
of the device, and thus choose  $r$ small enough so that the angular
divergence there fits the angular acceptance of the shaping device:
\begin{equation}
\frac{h}{r}>\frac{\Delta v}{\upsilon \sqrt{\varepsilon}}~,\quad
r<r_{\max} =\frac{\upsilon \sqrt{\varepsilon h}}{\sqrt{2\gb}}
=\sqrt{\varepsilon h h_{\max }}~.
\end{equation}
To write these relations, we have neglected the effect of gravity on
the short distance $r$ and used the value in \eqref{horizontal} of
the root-mean-square (rms) dispersion of horizontal velocity.

We then consider the role of the finite radius  $R$ of the disk,
using the following classical arguments. We want to produce an
efficient loss of atoms having too large velocities with respect to
the designed angular acceptance of the shaping velocity, and thus
choose $R$ large enough so those atoms efficiently touch the top
disk. Saying that they touch it at least once, this implies that the
time $T$ they spend in the zone between the disks is about two times
larger than the time $t_h=\sqrt{2h/\gb}$ corresponding to a free
fall on a height $h$:
\begin{align}
&T=\frac{R}{\upsilon \sqrt{\varepsilon}}>2t_h =
2\sqrt{\frac{2h}{\gb}}\notag\\
& \quad \Rightarrow \quad R>{R}_{\min }= \frac{4\upsilon
\sqrt{{\varepsilon h}}}{\sqrt{2\gb}}=4r_{\max }~. \label{timeT}
\end{align}
Again, we have used the dispersion \eqref{horizontal} of horizontal
velocity to calculate the time $T$ spent in the shaping device for
an atom with the rms velocity. Of course, the time  $T$ depends on
the actual horizontal velocity (not its rms value) so that a value
larger than that calculated in \eqref{timeT} is required to produce
an effective shaping for the whole distribution.

We want also to stress that the time $T$ appears as a systematical
delay in the free fall timing experiment so that its knowledge is
crucial for accuracy. Here the fact that annihilation event
detectors are position-sensitive is important. Measuring the
horizontal distance $L$ between the initial spot and the detection
point indeed gives the actual horizontal velocity of the atom
$L/T_{\tot}$ with $T_{\tot}$ the time between escape from the trap
and annihilation on the detector and allows one to correct the
timing measurement for the time spent in the shaping device $T=R
T_{\tot}/L$.

At the exit of the shaping device, the height lies in the interval
$\left[H,H+h\right]$ while the vertical velocity lies in the
interval $\left[-\Delta v,+\Delta v\right]$. As discussed in the
next section, this affects the resolution of the timing measurement
in the same manner as the dispersion of velocities did affect the
free fall measurement discussed in section \ref{spreading}. In order
to optimize the various parameters, in particular the value of the
radius $R$, we have to simulate the whole experiment, that is the photo-detachment, the
passage through the shaping device, the free fall from its output
slit to the detection plate, the timing of annihilation events, and
the correction from the time spent in the device. In the present
paper, we use simpler arguments to estimate the resulting accuracy
of the measurement.

\section{Estimation of statistical uncertainty}
\label{uncertainty}

At this point, we have all information needed to give a simple
estimation of the statistical accuracy in this experiment. To this
aim, we use the analogy with the free fall timing measurement to
write the relative spread of the free fall times as (compare with
\eqref{deltat}):
\begin{align}
 \frac{\Delta t}{t_H}=\sqrt{\alpha{\left(\frac{h}{2H}\right)}^{2}
 +\beta {\left(\frac{\Delta v}{v_H}\right)}^{2}}
\end{align}
 $\alpha $ and  $\beta $ are dimensionless numbers smaller than unity
describing the shapes of position and velocity distributions at the
output slit of the shaping device. For simplicity, we have supposed
that these distributions are uncorrelated and we have considered
that the correction for the time $T$ spent in the shaper has been
done. As $\Delta v=\sqrt{2\gb h}$ and $v_H=\sqrt{2\gb H}$ with
$h{\ll}H$, it follows that the relative spread $\left(\Delta
t/t_H\right)$ is dominated by the effect of velocity dispersion and
can be written as:
\begin{align}\label{deltatvel}
\frac{\Delta t}{t_H}\approx\sqrt{\frac{{\beta h}}{H}}~.
\end{align}
This corresponds to an accuracy $\left(\Delta \gb/\gb\right) \approx
2\sqrt{{\beta h}/H}$ for each detection of an annihilation event. We
then obtain the resolution after one month of measurement, taking
into account that the number of events is reduced by the velocity
selection (compare with \eqref{NNtot}):
\begin{align}\label{improvedaccuracy}
 \frac{\Delta\gb}{\gb\sqrt{N}}
 =2\sqrt{\frac{\beta h}{H}}\sqrt{\frac{\upsilon \sqrt{2\pi }} {N_\tot \Delta v}}
 =2 \! \left( \frac{{\pi h}{\beta }^{2}{\upsilon }^{2}}
 {\gb{H}^{2}N_{\tot }^{2}} \right)^{1/4}.
\end{align}

It is instructive to compare this resolution with the analogous
result obtained without the velocity selection mechanism. The
improvement is described by the ratio of \eqref{improvedaccuracy} to
\eqref{accuracy}:
\begin{align}\label{improvement}
2 \left( \frac{{\pi h}{\beta }^{2}{\upsilon }^{2}}{\gb{H}^{2}N_{\tot
}^{2}} \right)^{1/4}
 \left( \frac{2\upsilon}{v_H\sqrt{N_{\tot }}} \right)^{-1}
=\left( \frac{2{\pi h}{\beta }^{2}}{h_{\max }} \right)^{1/4}~.
\end{align}
The best accuracy is therefore achieved for smaller slit sizes. We
take the value $H=0.3$~m chosen in the current design of GBAR, the
worst case of $\beta=1$ and a velocity dispersion
$\upsilon=$~0.44~m/s and discuss three cases corresponding to
decreasing values of $h$:
\begin{enumerate}

\item Equation \eqref{improvement} shows that $h$ should be smaller
than $\approx h_{\max }/2\pi$ for the shaping device to improve the
resolution of the experiment. We choose as an example $h=1$~mm, so
that the statistics is $N\approx 3.3\times10^3$. The opening radius
has to be smaller than $r_{\max }\simeq 3.2 \sqrt{\varepsilon}$~mm
and the disk radius should be larger than $R_{\min }\simeq 13
\sqrt{\varepsilon}$~mm. The statistical accuracy is then
$\Delta\gb/\gb\sqrt{N} \approx 2.0\times10^{-3}$. Note that for a
conducting mirror and a maximal vertical velocity $\sqrt{2 g
h}\approx0.14$~m/s, the reflection probability for an atom is $78\%$
\cite{dufour2013}. To simultaneously improve the resolution and
reduce losses from annihilation on the bottom mirror, we move to
smaller values of $h$.

\item For $h<50$~$\mu$m, the atom flux through the slit can no longer
be evaluated from classical arguments and the quantum behavior of
$\Hb$ in the slit between the disks has to be taken into account
\cite{nesvizhevsky2002,nesvizhevsky2003,nesvizhevsky2005}. At
the boundary $h=50$~$\mu$m, the statistics is $N \approx
7.3\times10^2$ and the statistical accuracy is $\Delta
\gb/\gb\sqrt{N} \approx 1.0{\times}{10}^{-3}$. The opening radius
has to be smaller than $r_{\max }\simeq 0.7 \sqrt{\varepsilon}$~mm.
Note that the reflection probability for an atom with the maximal
velocity $\sqrt{2 g h}=3.1\times10^{-2}$~m/s is 94\% for a perfect
mirror.

\item For $h< 20~\mu$m, only atoms in the lowest quantum state
can pass through the slit. The reflection probability approaches
unity in this case which also corresponds to the highest accuracy
for the free fall timing measurement. This quantum limit is analyzed
in sections \ref{gravqstates} and \ref{freefallground}.
\end{enumerate}

The first and second cases provide more comfortable conditions for
merging the proposed shaping device and the Paul trap, as well as
better statistics. In this discussion, we have disregarded several
factors which may decrease statistics (annihilation of $\Hb$ in the
bottom disk, non-perfect merging of the angular acceptance of the
optical device and the incoming beam of  $\Hb$, quantum reflection
of $\Hb$ from the reference plate, etc.). These factors have to be
evaluated at a later stage.

\section{Quantum mechanical description}
\label{qmech}

We now perform a quantum-mechanical description of the experiment,
which will turn out to reproduce the main features and estimations
of the quasi-classical treatment given above.

\subsection{Free fall of a wave-packet}

We consider the free fall of a pre-formed quantum wave-packet of
$\Hb$ in the Earth's gravity field, and estimate the accuracy of the
corresponding time-of-fall measurement. We know that the initial
state ${\Psi }_{0}\left(z\right)$ of the wave-packet is a Gaussian
function centered in the vertical direction $z$ around the height
$H$ of the center of the trap, with the vertical position dispersion given by
\eqref{dispersions}:
\begin{align}
\Psi_0(z)= \left(\frac{m\omega}{\hbar \pi}\right)^{1/4}
\exp\left(-\frac{m \omega}{2\hbar}\left(z-H\right)^2\right)~.
\label{initialwave}
\end{align}
This wave-function is calculated prior to the release, at a time
where the gravity is compensated by the trap. After the
photo-detachment event, the atom is suddenly released and its state
is modified by the free fall in the gravity field.

This evolution is given by the propagation equation:
\begin{align}
\label{propagation}
\Psi\left(z,t\right)=
\int_{-\infty}^\infty G\left(z,z',t\right) \Psi_0\left(z'\right) \dd z'
\end{align}
where $t$ is the free fall time and $G$ the propagator:
\begin{align}
G\left(z,z',t\right)=&\sqrt{\frac{m}{2 i \pi \hbar t}}
\exp\left[\frac{im}{2\hbar t}\left(z-z'
+\frac{\gb t^2}{2}\right)^2\right]\notag \\
&\quad\times\exp \left[ \frac{m\gb zt+\frac16 m \gb^2 t^3}{i\hbar}
\right]~.\label{propagator}
\end{align}
Integrating \eqref{propagation} for the initially Gaussian
wave-packet \eqref{initialwave}, one gets :
\begin{align}
\Psi\left(z,t\right)&= \left(\frac{m\omega}{\hbar \pi (1+i\omega
t)^2}\right)^{1/4} \!\!\!\exp \left[ \frac{m\gb zt+\frac16 m \gb^2
t^3}{i\hbar} \right] \notag\\
&\times \exp\left[-\frac{m \omega}{2\hbar (1+i\omega t)}
\left(z-H+\frac{\gb t^2}{2}\right)^2\right]~.
\end{align}

\begin{figure}
\centering
   \includegraphics[width=8cm]{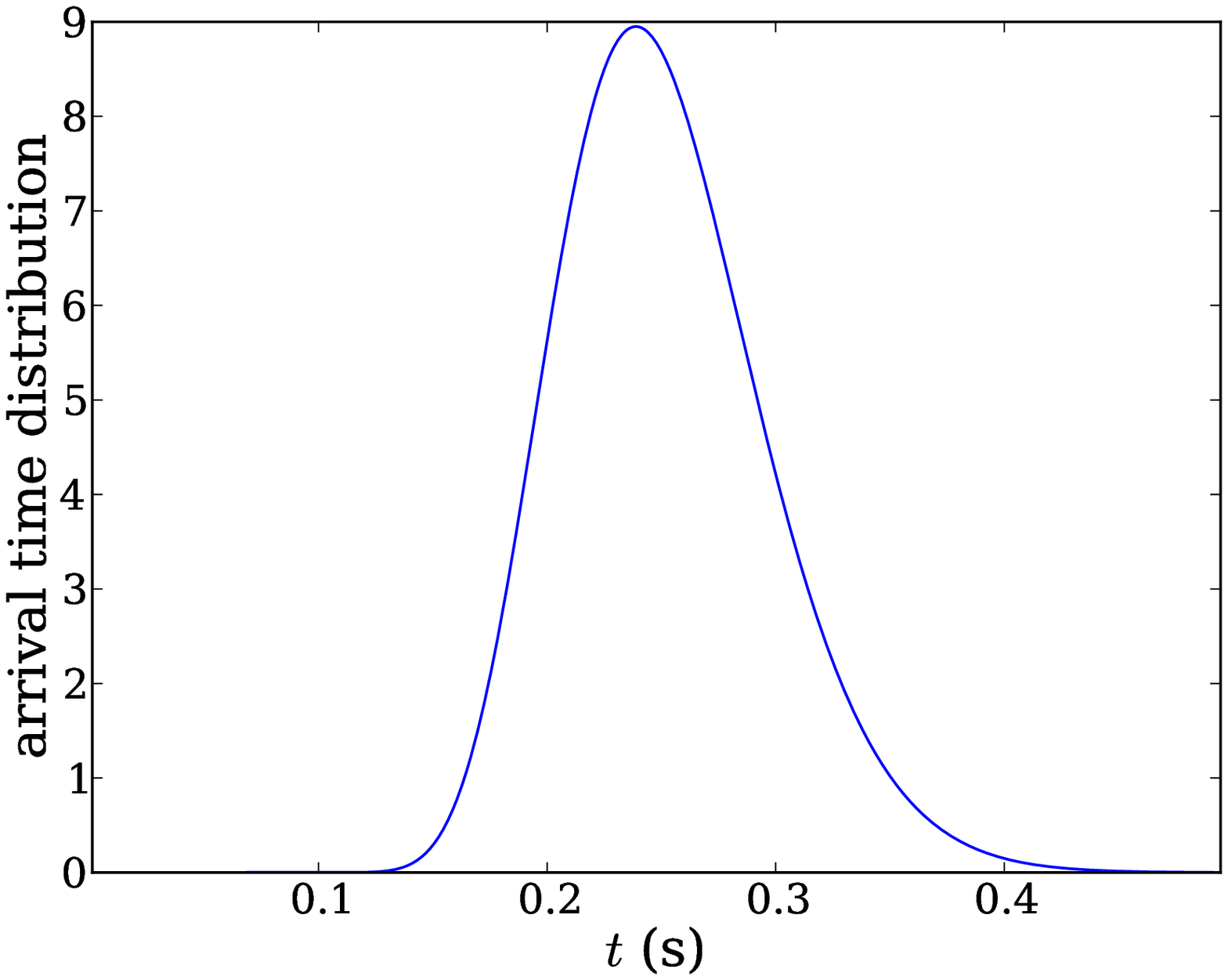}

   \includegraphics[width=8cm]{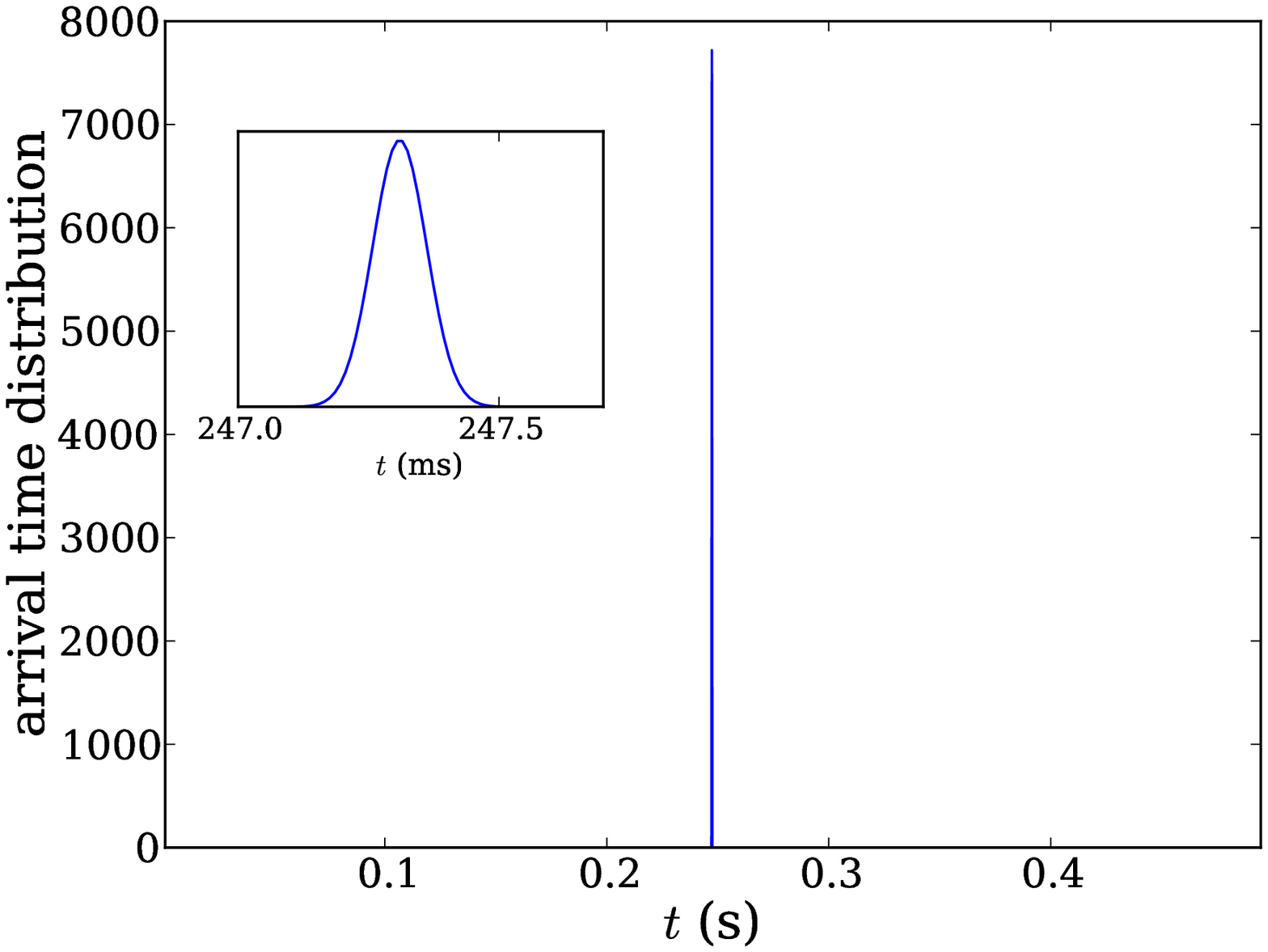}
\caption{\label{arrivaltime} The distribution of annihilation times
of $\Hb$ falling from the height $H=30$~cm; in the upper plot, the
initial state is a Gaussian of width $\zeta=70$~nm, a typical value
expected in the GBAR experiment; in the lower plot, it is a Gaussian
with the optimal width $\zeta_{opt}=88$~$\mu$m. For comparison, the
time scale is the same on both graphs. A zoom on the peak is shown
in the inset for the lower plot.}
\end{figure}

Assuming that all atoms annihilate instantaneously when they
touch the detection plate at $z=0$, we deduce that the distribution
for annihilation times is given by the flux $\mathcal{F}(t)$ of
atoms passing through the plane at height $z=0$, that is also the
opposite of the current (downward velocities have negative values):
\begin{align}
\mathcal{F}(t)&=-j(0,t)=-\frac{\hbar}{m} \text{Im}\left( \Psi(0,t)^*
\frac{\partial}{\partial z} \Psi(0,t) \right)~, \notag\\
&=\sqrt{\frac{m\omega^5 t^2}{\hbar\pi(1+\omega^2 t^2)^3}} \left( H +
\frac{\gb t^2}{2} + \frac{\gb}{\omega^2} \right)
\notag\\
&\times\exp\left[-\frac{m\omega}{\hbar (1+\omega^2 t^2)} \left(
\frac{\gb t^2}{2} - H \right)^2 \right]~. \label{probabilitydistrib}
\end{align}
This probability distribution is shown in figure
\ref{arrivaltime} for an initially Gaussian wavepacket dropped from
30~cm, in the two cases of an initial size typically expected for
the GBAR expected (upper plot) and the optimal size discussed above
(lower plot).

The optimal case (lower plot) leads to an extremely narrow time
distribution, with a peak having the Gaussian shape deduced by
expanding at lowest order in $(t-t_H)$ the distribution
\eqref{probabilitydistrib} :
\begin{align}
\mathcal{F}(t) \underset{t\approx t_H}{\simeq} C
\exp\left[-\frac{(t-t_H)^2}{2 \Delta t^2} \right]~.
\end{align}
The width $\Delta t$ of the distribution agrees with the classical
result \eqref{deltatheisenberg} :
\begin{align}
 \Delta t= \sqrt{ \frac{\hbar (1+\omega^2 t_H^2)}{2 m\omega\gb^2 t_H^2} }
 = t_H \sqrt{{\left(\frac{\zeta}{2H}\right)}^{2}+{\left(\frac{\hbar }
 {2m v_H\zeta}\right)}^{2}}~.
\end{align}
The upper plot in figure \ref{arrivaltime}, which corresponds to the
typical numbers of the GBAR design, leads to a much broader
distribution and shows a deformed shape with respect to a Gaussian
distribution. As already discussed, this is a consequence of the
large dispersion of initial vertical velocities.

\subsection{Gravitational quantum states in the shaping device}
\label{gravqstates}

We come now to the discussion of the shaping device in the regime
where quantum gravitational states play an important role. The
wave-function of the atoms can thus be developed over the basis of
eigenstates $\Psi_n$ with energies $E_n$ in the gravity field, here
calculated above a perfectly reflecting mirror \cite{voronin2006},
\begin{align}
\Psi_n(z)=\frac{1}{\sqrt{l}}\frac{\Ai(z/l-\lambda_n)}{\Ai'(-\lambda_n)}
\quad,\quad E_n=mgl\lambda_n~.
\end{align}
The typical scale $l$ of gravitational quantum states is:
\begin{align}
l=\left( \frac{\hbar^2}{2 m^2 \gb} \right)^{1/3} \approx
5.9~\mu\mathrm{m}~.
\end{align}
and the quantized energy levels are determined by the zeros of the
Airy function $\Ai$:
\begin{align}
\label{airyfunc}
&\Ai(-\lambda_n)=0 \\
&\lambda_1\approx 2.34 \;,\;\lambda_2\approx4.09 \;,\;
\lambda_3\approx5.52 \;,\; \ldots \notag
\end{align}
The high-$n$ states are given by the asymptotic law
\begin{align}
&\lambda_n\underset{n\to\infty}{\approx}\left( \frac{3\pi}{2}\left(
n-\frac{1}{4} \right) \right)^{2/3} \label{asymptotic}
\end{align}

Selectivity of the shaping device is based on the sharp dependence
of the transmission of eigenstates $\Psi_n$ versus the height $h$ of
the slit. The detailed formalism in \cite{voronin2006} leads to a
propagation through the device described by the following
propagator:
\begin{align}
K(z,z',t)=\sum_n \Psi_n(z) \Psi_n(z') \exp\left[ \frac{(E_n-i\Gamma_n)
t} {i\hbar} \right]~. \label{propagationGamma}
\end{align}
The width $\Gamma_n$ of level $n$ becomes large for high values of
$n$ \cite{voronin2006}, as explained by the following qualitative
interpretation. When the spatial dispersion $l\lambda_n$ of the
state $\Psi_n$ is smaller than the slit size $h$, the overlap with
the absorber is small and the atom has a high probability to pass
through the device ($\Gamma_n$ small). On the other hand, when
$l\lambda_n$ is larger than $h$, the overlap of the wave-function
with the absorber is significant and atoms have a high probability
to be absorbed ($\Gamma_n$ large).

\begin{figure}
\centering
     \includegraphics[width=8cm]{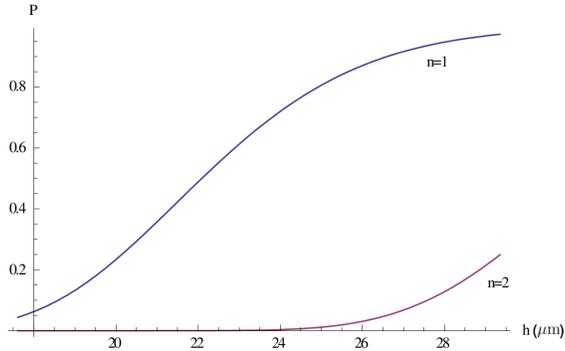}
\caption{\label{transmission} Transmission of first n=1 and second
n=2 gravitational states through a shaping device with a length of
5~cm.}
\end{figure}

As a quantitative illustration, figure \ref{transmission} shows the
probability of transmission for atoms in the two lowest
gravitational states $\Psi_1$ and $\Psi_2$ when the length of the
shaping device is $R-r=5$~cm and the roughness amplitude of the top
absorber is 1~$\mu$m. A slit size $h=24$~$\mu$m provides 72\%
transmission probability for the first state but only 0.3\% for the
second state. This implies that a nearly pure ground state or a
superposition of a few lowest gravitational states can be prepared
by a suitable choice of the parameters of the shaping device.

\subsection{Free fall experiment after the velocity shaping}
\label{freefallground}

The output of the velocity shaping device is a superposition of
gravitational quantum states $\Psi_n$, determined by the propagator
\eqref{propagationGamma} calculated for a time $t=(R-r)/v_\hor$ for
an atomic horizontal velocity $v_\hor$. This shaped superposition
then falls freely to the detection plate so that the time
distribution of annihilation events depends on the properties of the
shaped state. We stress again at this point that this supposes that
the time $R/v_\hor$ spent in the shaping device, and before its
entrance, is corrected in the data analysis, $v_\hor$ being deduced
from the position of the annihilation event.

The spatial and velocity dispersions of the state $\Psi_n$ can be
expressed in terms of $\lambda_n$ \cite{robinett2010}:
\begin{align}
&\Delta z_n=\frac{2l\lambda_n}{3 \sqrt{5}} &\Delta
v_n=\frac{\hbar}{ml}\sqrt{\frac{\lambda_n}{3}}
\end{align}
In contrast with the case of Gaussian wave-packets discussed above,
these dispersions do not reach the minimum in the Heisenberg
inequality. Furthermore, $\Delta v_n$ and $\Delta z_n$ increase
simultaneously as functions of $n$. The dispersion of the
annihilation time (after correction of the time spent in the device)
is thus given for the state $\Psi_n$ by \eqref{deltat} with
$\zeta,\upsilon$ replaced by $\Delta z_n,\Delta v_n$ :
\begin{align}
\frac{\Delta t}{t_H} = \sqrt{\frac{l^2 \lambda_n^2}{45 H^2}+\frac{l
\lambda_n}{3H}}{\approx} \sqrt{\frac{\lambda_n l}{3H}}
\label{simpleestim}
\end{align}
As $l\lambda_n\sim h\ll H$, the initial velocity spread still
dominates the uncertainty on the annihilation time. It follows that
the dispersion of these times is determined by $\Delta v_n$ and
scales as $\sqrt{\lambda_n}$.

In order to get an estimate of the dispersions, we suppose
that the state in the shaper is an incoherent superposition of the
quantum states which fit in the slit. It follows from the arguments
in the preceding section that the quantum states which fit in the
slit correspond to
\begin{equation}
n\leq n_{\max}~,\quad l\lambda_{n_{\max}}\approx h~.
\end{equation}
We then deduce the dispersion of annihilation times as
\begin{equation}
\frac{\Delta t}{t_H} = \sqrt{ \sum_{n} \pi_n \frac{l\lambda_n
}{3H}}~,
\end{equation}
where $\pi_n$ is the population in the state $\Psi_n$. As the slit
size is small compared with the incoming wave-function size, we
expect that the states are equally populated among the fitting
gravitational quantum states, so that $\pi_n\approx1/{n_{\max}}$ for
$n\leq n_{\max}$, $\pi_n\approx0$ otherwise. In the quasi-classical
limit where $n_{\max}\gg1$, we can use the asymptotic expression
\eqref{asymptotic} for $\lambda_n$ and replace the sum by an
integral to find:
\begin{equation}
\frac{\Delta t}{t_H} \approx \sqrt{\frac{ h}{5H}}~.
\end{equation}
This expression scales like the classical result \eqref{deltatvel}
with $\beta$ now specified to be $1/5$. 

The preceding argument disregards the coherence between the
components $\Psi_n$ in the superposition prepared by the shaping
device. This approximation can be justified qualitatively by
considering that the effects of coherence are washed out in the
averaging associated with free fall propagation as well as
horizontal velocity dispersion. However it cannot be considered as
exact, and it will have to be confirmed by more precise simulations,
to be published in forthcoming papers. 

\begin{figure}
\centering
     \includegraphics[width=8cm]{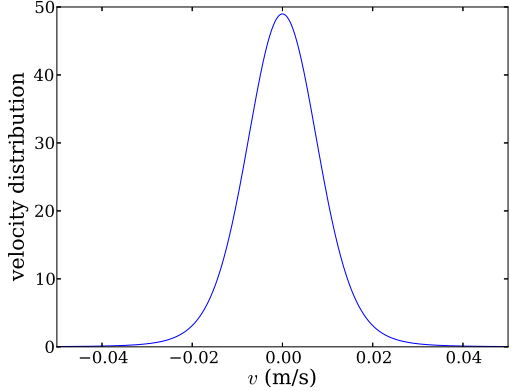}
\caption{\label{groundgrav} The velocity distribution in the ground
gravitational state $\Psi_1$.}
 \end{figure}

\begin{figure}
     \centering
     \includegraphics[width=8cm]{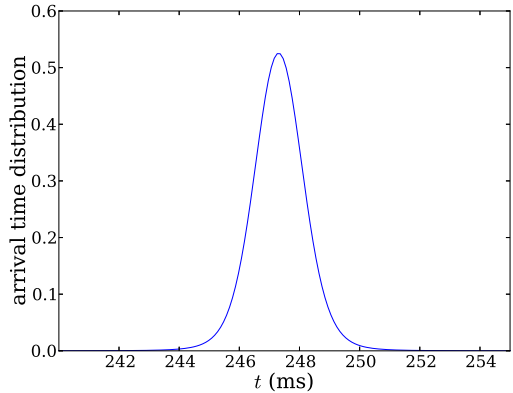}
\caption{\label{arrival_groundgrav}The distribution of arrival times
of $\Hb$ falling from the height of 30~cm, assuming that the initial
wave-packet has been shaped into the gravitational ground state
$\Psi_1$.}
\end{figure}

Exact quantum calculations can be performed for the special
case of an initial state for free fall prepared by the shaper as the
ground gravitational state $\Psi_1$. The initial velocity
distribution, shown in figure \ref{groundgrav}, has a width $\Delta
v \approx 9.5$~mm/s. This is 30 times larger than the optimal
velocity spread $\upsilon_{opt}\approx 0.36$~mm/s, but two orders of
magnitude smaller than the initial velocity spread in the GBAR
experiment. The exact quantum evolution of this initial wave-packet
is then obtained by integrating the propagation equations
(\ref{propagation}-\ref{propagator}). The annihilation time
distribution calculated in this manner is shown in figure
\ref{arrival_groundgrav}. Its spread is in excellent agreement with
the prediction $\Delta t = t_H \sqrt{{l\lambda_1}/{3H}} \simeq
0.97$~ms deduced from \eqref{simpleestim}. As a comparison, this
spread was of the order of 45~ms for the free fall measurement
performed without velocity shaping. The improvement reflects the
velocity selection by the shaping device, which is only partly
balanced by the degradation of the statistics (as discussed above).

\section{Estimation of systematic effects}
\label{systematics}

{For our proposal to be useful as an improved option of the GBAR
measurement, one must ensure that there are no large systematic
uncertainties which could contribute at a level comparable to the
estimated statistical uncertainty of  ${\ 10}^{-3}$.}

We first examine the additional velocity dispersion caused by the photo-detachment recoil.
As discussed in \cite{debu2012},  the vertical velocity dispersion due to the absorption of the photon and the positron emission can be kept small ($\sim 0.5$~m/s) by using a horizontal polarized laser beam with an energy tuned at around $\Delta E\approx10$~$\mu$eV $\approx 0.1$~cm$^{-1}$ above threshold. The photo-detachment cross section near threshold follows the Wigner law and can be estimated by using the available information in the literature  to be $\sigma=6.8\times10^{-26}(\Delta E/1 \text{ cm}^{-1})^{3/2}\approx 2\times10^{-27}$~m$^2$ \cite{harms1997,broad1976,lykke1991,blondel}. With a $P=1$~W laser beam tuned close the threshold energy $E_T=6083$~cm$^{-1}=0.76$~eV focused on an area $A=10$~$\mu$m $\times 10$~$\mu$m covering the Paul trap center, the photo-detachment rate is $R=\sigma P/A E_T=130$~s$^{-1}$.

In GBAR, antihydrogen ions can be produced only every 110 s, the ejection period of the antiproton decelerator at CERN. This time is sufficient to photo-detach the excess positron with high efficiency. The method is to illuminate the ion during a short enough time so as to define the start time with high precision, at a low enough repetition rate so that in case of successful photo-detachment, the free fall is completed before the next laser shot.
For example, since the free fall time on 30 cm is only 250 ms, laser shots of 100 $\mu$s duration at a repetition rate of 2 Hz during 100 s allows the start time to be known with enough precision ($4\times 10^{-4}$), it also avoids ambiguity on identifying the successful shot, and leads to a photo-detachment efficiency larger than of 90 \%. 

Since the velocity dispersion induced by the atomic recoil is of the same order as that from the confinement in the Paul trap, one would not gain by trying to get closer to the optimal cloud size. Finally, this effect is equivalent to a slightly warmer antihydrogen cloud, which changes the effective value of the frequency $\omega$ to be used in the calculations, without affecting the principle of the method.

A careful analysis of other systematic effects has to be performed in the
future, in particular for the following list of possibilities:

\begin{enumerate}[1)]
\item Uncertainty of shaping/measuring the distribution of vertical
velocity components of $\Hb$ within the range of acceptance of the
two-disk system;
\item Finite positioning and timing resolution for the detection
of annihilation events;
\item Accuracy and reliability for the correction for the time spent
in the shaping device;
\item Diffraction of atoms on the mirror edges;
\item Residual electromagnetic effects, and in particular patch effect
on mirror surfaces;
\item Defects of mechanical alignments, such as inclinations of the
disks and detection plate;
\item Finite precision of production and adjustment of optical elements;
\item Vibrations able to cause parasitic transitions between gravitational
quantum states.
\end{enumerate}
Monte-Carlo simulations of the experiment are underway;  they take  into account
photo-detachment, coupling of the shaping device with the Paul trap and detector vessel, as well as points 1) and 2).
For most of these systematic effects, one may also rely on the experience
accumulated in experiments with UCNs
\cite{nesvizhevsky2002,nesvizhevsky2010b,nesvizhevsky2003,nesvizhevsky2005}.
We note that the main systematic uncertainties (in particular 1) are
proportional to the ratio $h/H$, and thus decrease strongly when
slit heights are decreased. We therefore think that the control of
these systematic effects will be improved at small slit heights.

\section{Conclusion}

In this paper, we have proposed a new method for shaping vertical
velocities of antihydrogen atoms in the timing experiment to be
performed by the GBAR collaboration \cite{gbar}. We have given first
estimations of the corresponding statistic uncertainties and listed
possible systematic effects. The conclusion of these preliminary
estimations, to be confirmed by further analysis, is that the
accuracy in the measurement of the free fall acceleration $\gb$ of
$\Hb$ atoms could be pushed below 10$^{-3}$ in realistic
experimental conditions.

Statistical uncertainties in the experiment are improved for smaller
slit heights, which lead to better defined vertical velocities of
$\Hb$. This means that a better selection of the range of vertical
velocities overweighs the loss in statistics. Systematical
uncertainties are expected to decrease even more dramatically for
smaller heights of the slit between the two disks in the proposed
experimental design. In the optimum experiment where atomic
wave-packet is shaped to the lowest quantum state, the effective
temperature corresponding to the vertical motion of $\Hb$ is as low
as 10~nK.

These preliminary estimations have to be confirmed by more
complete simulations. We are currently working to develop a fully
quantum treatment of the shaping device as well as a complete
Monte-Carlo simulations.

Let us also mention that an even better accuracy could in principle
be obtained by studying interference effects in the time-of-arrival
distribution of a coherent superposition of a few lowest-lying
gravitational quantum states \cite{voronin2011,voronin2012}.

\section*{Acknowledgements}
The authors thank the ESF Research Networking Programme CASIMIR
(\href{http://casimir-network.org}{casimir-network.org}), the GRANIT collaboration and the GBAR collaboration (\href{http://gbar.in2p3.fr}{gbar.in2p3.fr}) for
providing excellent possibilities for discussions and exchange.

\bibliographystyle{unsrt}
\bibliography{biblio}

\begin{thebibliography}{10}

\bibitem{kellerbauer2008}
A.~Kellerbauer, M.~Amoretti, {A.S.} Belov, G.~Bonomi, I.~Boscolo, {R.S.} Brusa,
  M.~B\"{u}chner, {V.M.} Byakov, L.~Cabaret, C.~Canali, C.~Carraro,
  F.~Castelli, S.~Cialdi, M.~de~Combarieu, D.~Comparat, G.~Consolati,
  N.~Djourelov, M.~Doser, G.~Drobychev, A.~Dupasquier, G.~Ferrari, P.~Forget,
  L.~Formaro, A.~Gervasini, {M.G.} Giammarchi, {S.N.} Gninenko, G.~Gribakin,
  {S.D.} Hogan, M.~Jacquey, V.~Lagomarsino, G.~Manuzio, S.~Mariazzi, {V.A.}
  Matveev, {J.O.} Meier, F.~Merkt, P.~Nedelec, {M.K.} Oberthaler, P.~Pari,
  M.~Prevedelli, F.~Quasso, A.~Rotondi, D.~Sillou, {S.V.} Stepanov, {H.H.}
  Stroke, G.~Testera, {G.M.} Tino, G.~Tr\'{e}nec, A.~Vairo, J.~Vigu\'{e},
  H.~Walters, U.~Warring, S.~Zavatarelli, and {D.S.} Zvezhinskij.
\newblock Proposed antimatter gravity measurement with an antihydrogen beam.
\newblock {\em Nuclear Instruments and Methods in Physics Research Section B:
  Beam Interactions with Materials and Atoms}, 266(3):351--356, February 2008.

\bibitem{alpha}
The~ALPHA Collaboration and A.~E. Charman.
\newblock Description and first application of a new technique to measure the
  gravitational mass of antihydrogen.
\newblock {\em Nature Communications}, 4:1785, April 2013.

\bibitem{atrap}
G~Gabrielse.
\newblock The production and study of cold antihydrogen.
\newblock Technical Report CERN-SPSC-2010-006.SPSC-SR-057, CERN, Geneva, Jan
  2010.

\bibitem{gbar}
G~Chardin, P~Grandemange, D~Lunney, V~Manea, A~Badertscher, P~Crivelli,
  A~Curioni, A~Marchionni, B~Rossi, A~Rubbia, V~Nesvizhevsky, {P-A} Hervieux,
  G~Manfredi, P~Comini, P~Debu, P~Dupr\'{e}, L~Liszkay, B~Mansouli\'{e},
  P~P\'{e}rez, {J-M} Rey, N~Ruiz, Y~Sacquin, A~Voronin, F~Biraben, P~Clad\'{e},
  A~Douillet, A~G\'{e}rardin, S~Guellati, L~Hilico, P~Indelicato, A~Lambrecht,
  R~Gu\'{e}rout, {J-P} Karr, F~Nez, S~Reynaud, {V-Q} Tran, A~Mohri, Y~Yamazaki,
  M~Charlton, S~Eriksson, N~Madsen, {D-P} van~der Werf, N~Kuroda, H~Torii, and
  Y~Nagashima.
\newblock Proposal to measure the gravitational behaviour of antihydrogen at
  rest.
\newblock Technical Report CERN-SPSC-2011-029.SPSC-P-342, CERN, Geneva, Sep
  2011.

\bibitem{walz2004}
Jochen Walz and Theodor~W. H\"{a}nsch.
\newblock A proposal to measure antimatter gravity using ultracold antihydrogen
  atoms.
\newblock {\em General Relativity and Gravitation}, 36(3):561--570, March 2004.

\bibitem{dufour2013}
G.~Dufour, A.~G\'{e}rardin, R.~Gu\'{e}rout, A.~Lambrecht, V.~V. Nesvizhevsky,
  S.~Reynaud, and A.~Yu. Voronin.
\newblock Quantum reflection of antihydrogen from the casimir potential above
  matter slabs.
\newblock {\em Physical Review A}, 87(1):012901, January 2013.

\bibitem{nesvizhevsky2002}
Valery~V. Nesvizhevsky, Hans~G. B\"{o}rner, Alexander~K. Petukhov, Hartmut
  Abele, Stefan Bae{\ss}ler, Frank~J. Rue{\ss}, Thilo St\"{o}ferle, Alexander
  Westphal, Alexei~M. Gagarski, Guennady~A. Petrov, and Alexander~V. Strelkov.
\newblock Quantum states of neutrons in the earth's gravitational field.
\newblock {\em Nature}, 415(6869):297--299, January 2002.

\bibitem{nesvizhevsky2010a}
Valery~V. Nesvizhevsky, Alexei~Yu Voronin, Robert Cubitt, and Konstantin~V.
  Protasov.
\newblock Neutron whispering gallery.
\newblock {\em Nature Physics}, 6(2):114--117, February 2010.

\bibitem{nesvizhevsky2010b}
Valerii~V. Nesvizhevsky.
\newblock Near-surface quantum states of neutrons in the gravitational and
  centrifugal potentials.
\newblock {\em {Physics-Uspekhi}}, 53(7):645, October 2010.

\bibitem{ignatovich1990}
Vladimir~Kazimirovich Ignatovich.
\newblock {\em The physics of ultracold neutrons}.
\newblock Oxford series on neutron scattering in condensed matter ; Oxford
  science publications , Oxford University Press. Clarendon Press, Oxford,
  {Royaume-Uni}, 1990.

\bibitem{golub1991}
Robert Golub, David~J Richardson, and Lamoreaux ~.
\newblock {\em Ultra-cold neutrons}.
\newblock Adam Hilger, Bristol; Philadelphia, 1991.

\bibitem{voronin2011}
A.~Yu Voronin, P.~Froelich, and V.~V. Nesvizhevsky.
\newblock Gravitational quantum states of antihydrogen.
\newblock {\em Physical Review A}, 83(3):032903, 2011.

\bibitem{voronin2012}
A.~Yu Voronin, V.~V. Nesvizhevsky, and S.~Reynaud.
\newblock Interference of the whispering gallery states of antihydrogen.
\newblock {\em J. Physics B}, 45:165007, 2012.

\bibitem{nesvizhevsky2003}
V.~V. Nesvizhevsky, H.~G. B\"{o}rner, A.~M. Gagarski, A.~K. Petoukhov, G.~A.
  Petrov, H.~Abele, S.~Bae{\ss}ler, G.~Divkovic, F.~J. Rue{\ss}, Th.
  St\"{o}ferle, A.~Westphal, A.~V. Strelkov, K.~V. Protasov, and A.~Yu.
  Voronin.
\newblock Measurement of quantum states of neutrons in the
  earth{\textquoteright}s gravitational field.
\newblock {\em Physical Review D}, 67(10):102002, May 2003.

\bibitem{nesvizhevsky2005}
V.~V. Nesvizhevsky, A.~K. Petukhov, H.~G. Borner, T.~A. Baranova, A.~M.
  Gagarski, G.~A. Petrov, K.~V. Protasov, A.~Y. Voronin, S.~Baessler, H.~Abele,
  A.~Westphal, and L.~Lucovac.
\newblock Study of the neutron quantum states in the gravity field.
\newblock {\em European Physical Journal C}, 40(4):479--491, April 2005.

\bibitem{nesvizhevsky2000}
{V.V} Nesvizhevsky, H~B\"{o}rner, {A.M} Gagarski, {G.A} Petrov, {A.K} Petukhov,
  H~Abele, S~B\"{a}{\ss}ler, T~St\"{o}ferle, and {S.M} Soloviev.
\newblock Search for quantum states of the neutron in a gravitational field:
  gravitational levels.
\newblock {\em Nuclear Instruments and Methods in Physics Research Section A:
  Accelerators, Spectrometers, Detectors and Associated Equipment},
  440(3):754--759, February 2000.

\bibitem{voronin2006}
A.~Yu. Voronin, H.~Abele, S.~Bae{\ss}ler, V.~V. Nesvizhevsky, A.~K. Petukhov,
  K.~V. Protasov, and A.~Westphal.
\newblock Quantum motion of a neutron in a waveguide in the gravitational
  field.
\newblock {\em Physical Review D}, 73(4):044029, February 2006.

\bibitem{meyerovich2006}
A.~E. Meyerovich and V.~V. Nesvizhevsky.
\newblock Gravitational quantum states of neutrons in a rough waveguide.
\newblock {\em Physical Review A}, 73(6):063616, June 2006.

\bibitem{adhikari2007}
R.~Adhikari, Y.~Cheng, A.~E. Meyerovich, and V.~V. Nesvizhevsky.
\newblock Quantum size effect and biased diffusion of gravitationally bound
  neutrons in a rough waveguide.
\newblock {\em Physical Review A}, 75(6):063613, June 2007.

\bibitem{escobar2011}
M.~Escobar and A.~E. Meyerovich.
\newblock Beams of gravitationally bound ultracold neutrons in rough
  waveguides.
\newblock {\em Physical Review A}, 83(3):033618, March 2011.

\bibitem{walther2012}
A.~Walther, F.~Ziesel, T.~Ruster, S.~T. Dawkins, K.~Ott, M.~Hettrich,
  K.~Singer, F.~{Schmidt-Kaler}, and U.~Poschinger.
\newblock Controlling fast transport of cold trapped ions.
\newblock {\em Physical Review Letters}, 109(8):080501, August 2012.

\bibitem{robinett2010}
R.~W. Robinett.
\newblock The stark effect in linear potentials.
\newblock {\em European Journal of Physics}, 31(1):1, January 2010.

\bibitem{debu2012}
Pascal Debu.
\newblock {GBAR}.
\newblock {\em Hyperfine Interactions}, 212(1-3):51--59, December 2012.

\bibitem{harms1997}
Oliver Harms, Michael Zehnpfennig, Victor Gomer, and Dieter Meschede.
\newblock Photodetachment spectroscopy of stored ions.
\newblock {\em Journal of Physics B: Atomic, Molecular and Optical Physics},
  30(17):3781, September 1997.

\bibitem{broad1976}
John~T. Broad and William~P. Reinhardt.
\newblock One- and two-electron photoejection from {H\textsuperscript{-}}: A
  multichannel j-matrix calculation.
\newblock {\em Physical Review A}, 14(6):2159--2173, December 1976.

\bibitem{lykke1991}
K.~R. Lykke, K.~K. Murray, and W.~C. Lineberger.
\newblock Threshold photodetachment of {H\textsuperscript{-}}.
\newblock {\em Physical Review A}, 43(11):6104--6107, June 1991.

\bibitem{blondel}
The formula given is a result of a compilation made by C. Blondel, private
  communication (2012).

\end{thebibliography}

 \end{document}